\input{aipcheck}
\documentclass[ ,final ,numberedheadings ]
  {aipproc}
\usepackage{amsmath}

\layoutstyle{8x11single}

\begin{document}

\title{The intriguing evolutionary history of the massive black hole X-ray binary 
M33 X-7.}


\author{Francesca Valsecchi}{
  address={Center for Interdisciplinary Exploration and Research in Astrophysics (CIERA) and Department of Physics and Astronomy, Northwestern University, 2145 Sheridan Road, Evanston, IL 60208, USA}
}

\author{Evert Glebbeek}{
  address={Department of Physics and Astronomy, McMaster University,  1280 Main Street West, Hamilton, Ontario, Canada L8S 4M1}
}

\author{Will M. Farr}{
  address={Center for Interdisciplinary Exploration and Research in Astrophysics (CIERA) and Department of Physics and Astronomy, Northwestern University, 2145 Sheridan Road, Evanston, IL 60208, USA}
}
\author{Tassos Fragos}{
  address={Center for Interdisciplinary Exploration and Research in Astrophysics (CIERA) and Department of Physics and Astronomy, Northwestern University, 2145 Sheridan Road, Evanston, IL 60208, USA}
  ,altaddress={Harvard-Smithsonian Center for Astrophysics, 60 Garden Street, Cambridge, MA 02138, USA}
}
\author{Bart Willems}{
  address={Center for Interdisciplinary Exploration and Research in Astrophysics (CIERA) and Department of Physics and Astronomy, Northwestern University, 2145 Sheridan Road, Evanston, IL 60208, USA}
}
\author{Jerome A. Orosz}{
  address={Department of Astronomy, San Diego State University, 5500 Campanile Drive, San Diego, CA 92182-1221, USA}
}
\author{Jifeng Liu}{
 address={National Astronomical Observatories, Chinese Academy of Sciences, Beijing 100012, China}
  ,altaddress={Harvard-Smithsonian Center for Astrophysics, 60 Garden Street, Cambridge, MA 02138, USA}
}
\author{Vassiliki Kalogera}{
  address={Center for Interdisciplinary Exploration and Research in Astrophysics (CIERA) and Department of Physics and Astronomy, Northwestern University, 2145 Sheridan Road, Evanston, IL 60208, USA}
}

\begin{abstract}
Black hole (BH) X-ray binaries (XRBs) are X-ray luminous binary systems comprising a BH accreting matter from a companion star. 
Understanding their origins sheds light on the still not well understood physics of BH formation. 
M33 X-7 hosts one of the most massive stellar-mass BH among all XRBs known to date, 
a 15.65$\, \rm{M}_\odot$ BH orbiting a 70$\, \rm{M}_\odot$ companion star in a 3.45$\,$day orbit. 
The high masses of the two components and the tight orbit relative to the large H-rich stellar component challenge our understanding of the typically invoked BH-XRBs formation channels. The measured underluminosity of the optical 
component further complicates the picture. A solution to the evolutionary history of this system that can
 account for all its observed properties has yet to be presented, and here we propose the first scenario
  that is consistent with the complete set of current observational constraints. In our model, M33 X-7 started
   its life hosting a 85-99$\, \rm{M}_\odot$ primary and a 28-32$\, \rm{M}_\odot$ companion in a Keplerian
    orbit of 2.8-3.1$\,$days. In order to form a BH of 15.65$\, \rm{M}_\odot$, the initially most massive
     component transferred part of its envelope to the companion star and lost the rest in a strong stellar wind. 
     During this dynamically stable mass transfer phase the companion accreted matter, to become the presently underluminous 70$\, \rm{M}_\odot$ star. 
\end{abstract}
\maketitle

%%%%%%%%%%%%%%%%%%%%%%%%%%%%%%%%%%%%%%%%%%%%
%% MAINMATTER
%%%%%%%%%%%%%%%%%%%%%%%%%%%%%%%%%%%%%%%%%%%%

\section{Introduction}
X-ray binaries (XRBs) are key systems for studying the physics of black holes (BH).
In fact, although solitary BHs do not emit electromagnetic radiation, they become detectable
 X-ray sources when they have a stellar companion transferring matter to them. 
M33 X-7 provides a unique physical
laboratory for the study of BHs, massive stars and XRBs.
This system harbors the heaviest star ever discovered in an XRB orbiting one of the most massive 
BHs found in this class of systems,  a 70$\, \rm{M}_\odot$ star orbiting a 15.65$\, \rm{M}_\odot$ BH every
 3.45$\,$days. 
The massive components and the tight orbit mark M33 X-7 as an evolutionary challenge. The search for a plausible evolutionary scenario is further complicated by the luminosity of the stellar component, which is lower than what evolutionary models predict for a single star of 70$\, \rm{M}_\odot$\cite{2007Nature}.
Our analysis aims to find a solution to the evolutionary history of M33 X-7 that can simultaneously explain  all its observed properties. 
\section{M33 X-7's Current Properties and Distance Uncertainty}
In Table\ref{obsProperties} we summarize some of the observed system's properties for a distance to M33 as adopted by the discovery team, and we discuss how some of these parameters would change if a different distance is considered. In fact, M33's distance is not well constrained by the literature, 
and various studies place it between 750-1017$\,$kpc \cite{UetAl2009, ScowcroftEtAl2009, 2007Nature, BonanosEtAl2006, SarajediniEtAl2006, CiardulloEtAl2004, GalletiEtAl2004, McConnachieEtAl2004, TiedeEtAl2004, KimEtAl2002, LeeEtAl2002, FreedmanEtAl2001, PierceEtAl2000, SarajediniEtAl2000}.
At the present time, the BH is accreting part of the matter the companion star is losing via stellar wind. In fact, given the extreme mass ratio, a phase of mass transfer (MT) through Roche-lobe (RL) overflow would be dynamically unstable and rapidly evolve into a merger of the two components.
\begin{table}[h]
\begin{tabular}{lrclr}
\hline
Parameter & Value&\phantom{a} & Parameter & Value\\ 
\hline
\smallskip
$M\rm_{BH}$ (M$_\odot$) &15.65 $\pm$ 1.45  & &$P$ (days) &3.45301 $\pm$ 0.00002 \\
\smallskip
$M\rm_{2}$ (M$_\odot$) & 70.0 $\pm$ 6.9&\phantom{a}& $e$ &0.0185 $\pm$ 0.0077 \\
\smallskip
Spectral Type & O7 III to O8 III & \phantom{a} & $i$ ($^\circ$)&74.6 $\pm$ 1.0\\
\smallskip
$T\rm_{eff}$ (K) &35000 $\pm$ 1000   &\phantom{a} &  $L\rm_{X} (10^{38}$ erg s$^{-1})$& 0.13 to 2.49\\
\smallskip
 Log($L\rm_{2}$/L$_\odot$)& 5.72 $\pm$ 0.07&\phantom{a}& $d$ (kpc)& 840 $\pm$ 20\\ 
\hline
\end{tabular}
\caption{Observed parameters for M33 X-7. The BH mass ($M_{\rm{BH}}$), its companion mass ($M_{\rm{2}}$), Spectral Type, effective temperature ($T_{\rm{eff}}$), and luminosity ($L_{\rm{2}}$), the orbital eccentricity ($e$), and inclination ($i$), and M33 distance ($d$), are taken from \citet{2007Nature}. \citet{Pietsch2006} determined the orbital period ($P$). The X-ray luminosity ($L_{\rm{X}}$) accounts for variations in the X-Ray flux over different observations \cite{ParmarEtAl2001, PietschEtAl2004, Pietsch2006, ShporerEtAl2007, 2007Nature, LiuEtAl2008}. If the full distance range of 750-1017$\,$kpc is adopted, via the ELC  \cite{Orosz2000} we calculate $M_{\rm{2}}\sim$ 55-103$\,$M$_\odot$, $M_{\rm{BH}}\sim$13.5-20$\,$M$_\odot$, $i\sim$77$^\circ$-71$^\circ$,  $\log(L_{\rm{2}}/L_\odot)\sim$5.62-5.89, and $L_{\rm{X}}$ between $\simeq$1$\times$10$^{37}$ and $\simeq$3$\times$10$^{38}$$\,$erg sec$^{-1}$.
}
\label{obsProperties}
\end{table}
\section{Modeling The System}
To investigate a solution to M33 X-7's history, we perform binary evolution calculations considering a variety of progenitor masses and initial orbital periods. 
Guided by the luminosity of the stellar component,  lower than what is expected from evolutionary models of single stars, we explore binary sequences where the BH companion accretes mass from the BH progenitor. 
Furthermore, given the short orbital period, we consider binary systems that start their life already in a tight orbit, hence undergoing a phase of MT during the Main Sequence (MS) phase of the primary component. 
The stellar evolution models are calculated with an up-to-date version of Eggleton's
stellar evolution code, STARS\cite{Eggleton1971, Eggleton1973, Polsetal1995, Eggleton2002}.
Given the spatial metallicity gradient of M33 \cite{UetAl2009}, we assume a metallicity 50\% of the solar value for all our models.
\subsection{Scan of the Parameter Space}
We evolve binary systems from the Zero-Age MS until the end of the primary's MS considering progenitor masses between 20-130$\,$M$_\odot$ and 10-100$\,$M$_\odot$ for the primary and the secondary, respectively, and initial orbital periods ranging from 1 to 10$\,$days.
We perform a first scan of the parameter space rejecting the sequences where the primary overfills its RL only after the end of its MS.  Furthermore, given the present high mass of the star, we also exclude the sequences where the secondary transfers mass back to the primary after having accreted from it. In our model, assuming quasi-conservative MT (with the only mass loss from the system being due to the stellar wind of both components), the primary transfers most of its H-rich envelope to the secondary, and becomes a  Wolf-Rayet star (WR). The stronger WR wind then interrupts the MT and blows away the remaining primary's envelope to expose its massive He core. 

At the end of the primary's MS, we use evolutionary models of single He stars (see Appendix A) to determine the mass lost from the system, and the consequent change in orbital period during the primary's core He burning phase, until collapse \footnote{The final stages of the primary's life  during and beyond carbon burning last too short to significantly change the system's parameters ($\sim$60$\,$yr for an initially $\simeq$25$\,$M$_\odot$ He star)\cite{TaurisHeuvel2003}}. 
We then reject the sequences where, at He exhaustion, the components masses are lower than the minimum observed values. 
Since no other episodes of RL overflow can have occurred from the primary's collapse, we evolve each secondary as a single star, and reject the sequences where the mass of the star does not fall within the observed range when the model matches the currently observed effective temperature and luminosity. After the primary's collapse, we study the evolution in time of the orbital separation, eccentricity, and spin of the stellar component accounting for torque between the binary components (following \citet{Belczynski2008}, and references therein\footnote{For the 2$^{nd}$ order tidal coefficient $E_2$ we use stellar models from Claret \cite{Claret2006} for masses of $\sim63\,$M$_\odot$ and $\sim79.5\,$M$_\odot$, and derive $E_2 = - 5.4566 - 7.37243\cdot t\rm_{MS}^{4.50562}$, where $t\rm_{MS}$ is time in units of the MS lifetime. The formula has only a very weak dependence on the initial mass for the range of interest.} ), changes in the stellar radius during the MS lifetime of the companion star, stellar wind mass loss \cite{EggletonBook2006}, orbital angular momentum loss due to gravitational radiation \cite{Junker1992}, and accretion from the companion's stellar wind onto the BH  \cite{BondiHoyle1944}. We consider the BH as a point mass. We explore a variety of initial orbital configurations, by scanning the parameter space made up of the kick magnitude ($V_{\rm{k}}$), orbital separation ($a_{\rm{postBH}}$), and eccentricity ($e_{\rm{postBH}}$), following \citet{Kalogera1996, WillemsEtAl2005}.
Specifically, we consider kicks between 0-1300 km/s, orbital separation between 0-100$\,$R$_\odot$, and eccentricities between 0 and 1. We assume an isotropic distribution for the direction of the kick. We interrupt the calculation when the orbital period crosses the observed value and the eccentricity of the orbit falls within the observed range.
Finally, of the sequences that fulfill all the above requirements, we reject the sequences where the star's radius at present is bigger than the distance from the center of the star to the L1 point \cite{SepinskyetalWithFred2007}. 
\subsection{Example of a Successful Evolutionary Sequence}
\begin{figure}
  \includegraphics[height=.42\textheight]{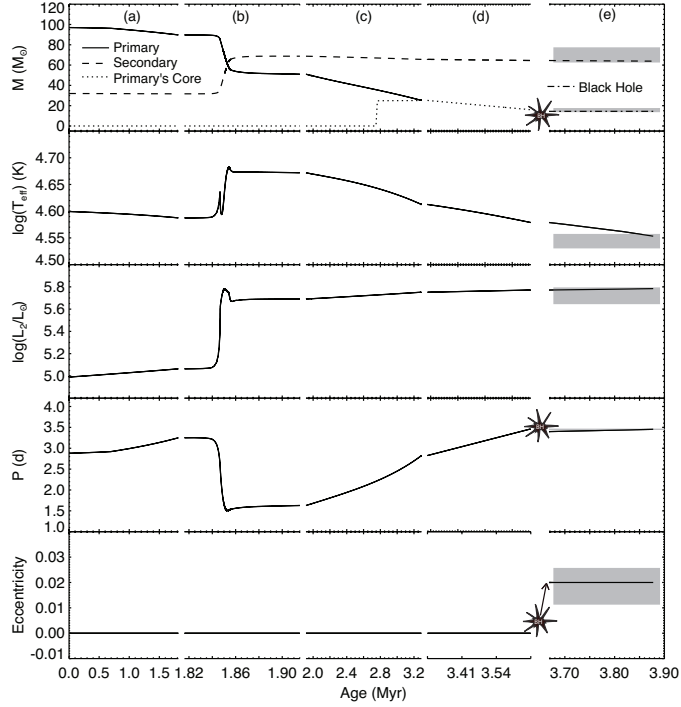}
  \caption{Evolution of the stellar and orbital parameters of M33 X-7. From the top: components masses ($M$), secondary's effective temperature [$\log(T_{\rm{eff}}$)] and luminosity [$\log(L_{\rm{2}}$)], orbital period ($P$), and eccentricity ($e$). From the left, the different evolutionary stages are: the beginning of the MS (a), the MT phase (b), the end of the MS (c), the core He burning phase for the primary (d), and the post BH formation phase (e) until the present time (note the non uniform x-axis).
The grey-shaded areas represent the observational constraints as reported in Table~\ref{obsProperties}. For the case shown, a kick of 120 km/s is imparted to the newly born BH.}
\label{totalEv}
\end{figure}

In Figure \ref{totalEv} we show the results for one of the successful evolutionary sequences, typical of M33 X-7's history. The progenitors comprise a primary and a secondary of $\simeq$97$\,$M$_\odot$ and $\simeq$32$\,$M$_\odot$, respectively, in an orbit of $\simeq$2.9$\,$days.
During the first $\simeq$1.8$\,$ Myr the evolution is driven by mass loss via stellar winds, causing an expansion of the orbit to $\simeq$3.25$\,$days.
While still on its MS the primary overfills its RL and begins MT onto the secondary. This stronger mode of mass loss brings the primary out of thermal equilibrium, and in response the star shrinks, recovering its thermal equilibrium while always maintaining hydrostatic equilibrium, and hence keeping the MT dynamical stable.
During the MT phase, while the more massive primary is transferring mass to the less massive secondary, the orbit shrinks, but when the secondary becomes the more massive component the orbit begins expanding \cite{Verbunt1993}. 
The primary transfers most of its H-rich envelope and becomes a WR star, and the strong WR wind ($\sim$2 to 3$\cdot\rm{10}^{-5}$ $\,$M$_\odot$yr$^{-1}$) eventually interrupts the MT.
During the $\simeq$99,000$\,$ years of conservative MT, the primary becomes a $\simeq$51$\,$M$_\odot$ WR star, while the original 32$\,$M$_\odot$ 
secondary becomes a massive $\simeq$69$\,$M$_\odot$ O-type star. 	
Once the WR wind sets in and the primary detaches, the wind blows away the remaining primary's envelope 
to expose the $\simeq$25$\,$M$_\odot$ He core. At the same time, the now more massive secondary is losing mass via its own wind at a lower rate ($\sim10^{-6}\,$M$_\odot$yr$^{-1}$). This mass loss causes the orbital period to further increase until the end of the primary's MS and throughout its core He burning phase.   
At this time the orbit is circular and the spin period of each star is expected to be synchronized with the orbital period. 
At primary's collapse, after $\simeq$3.7$\,$Myr,  M33 X-7 hosts a $\simeq$16$\,$M$_\odot$ 
evolved WR star, and a $\simeq$64.5$\,$M$_\odot$ O-star companion in a $\simeq$3.5$\,$day orbit.
Unable to support itself through further nuclear fusion, the WR star collapses into a BH, and 10\% of the rest mass energy is released as the BH's gravitational energy. Additionally, collapse asymmetries and associated neutrino emission $may$ impart a kick to the newly born BH, even with no baryonic mass ejection at collapse.
Both these effects modify the orbital configuration, slightly shrinking the orbit to $\simeq$3.4$\,$days, and inducing an eccentricity. For the remaining $\simeq$0.2$\,$Myr, the evolution of the system is driven by mass loss via the secondary's stellar wind, causing a further expansion of the orbit, and bringing the orbital period to the currently observed value. The fraction of this stellar wind accreted by the BH is too small to significantly influence the orbital evolution, but it is adequate to explain the observed X-Ray luminosity. At the present time, after $\simeq$3.9$\,$Myr, M33 X-7 comprises a BH of  $\simeq$14.4$\,$M$_\odot$
 and an underluminous O-star of  $\simeq$64$\,$M$_\odot$ in a slightly eccentric $\simeq$3.45$\,$day orbit. 
\begin{figure}[h]
  \includegraphics[height=.18\textheight]{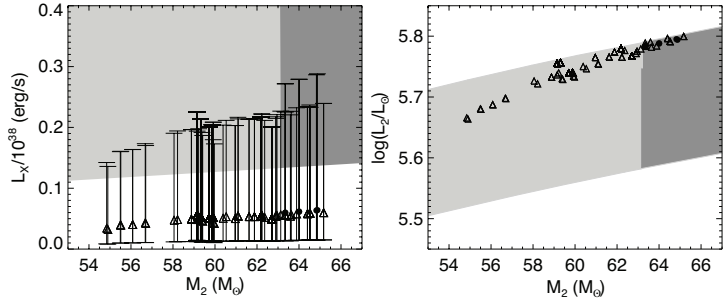}
  \caption{Current luminosity. The circles and triangles are the results of detailed binary star evolution calculations for all successful sequences for a M33 distance of 840 $\pm$ 20$\,$kpc, and of 750-1017$\,$kpc, respectively. The BH X-Ray luminosity (left), and the secondary's luminosity (right) are shown as a function of the secondary's mass at present. The dark and light grey shaded areas represent the observational constraints for a distance of 840 $\pm$ 20$\,$kpc, and of 750-1017$\,$kpc, respectively. $L_{\rm{X}}$ is calculated according to \citet{BondiHoyle1944}, and the error bars account for the uncertainties in the stellar wind parameters, and depict the highest and lowest $L_{\rm{X}}$ values. Some of the data points are omitted for clarity.}
\label{luminosities}
\end{figure}
\subsection{Parameter Space Left}
\begin{figure}[h]
  \includegraphics[height=.18\textheight]{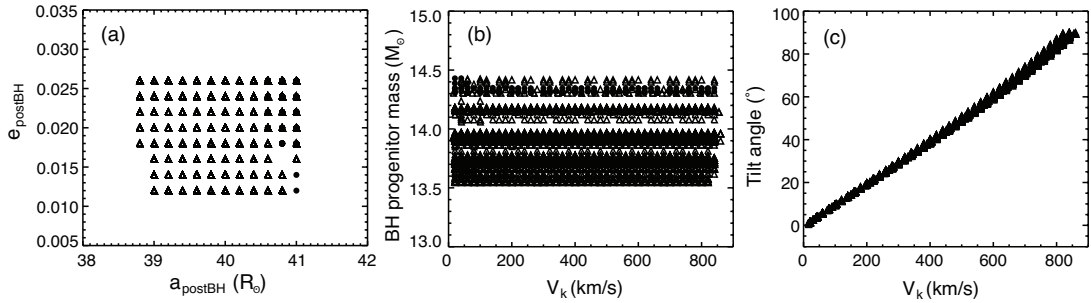}
  \caption{Orbital parameters at BH formation. The data points are as in Figure \ref{luminosities}. (a) eccentricity as a function of the orbital separation post-BH formation; (b) mass of the BH progenitor and (c) change in the orbital inclination at BH formation as a function of the kick magnitude. The BH progenitor mass accounts for the 10\% of rest mass energy released as the BH's gravitational energy at collapse. Given that the BH progenitor mass lies within the observed range for the BH mass, no baryonic mass is ejected at collapse. Some of the data points are omitted for clarity.}
\label{SNparams}
\end{figure}

All the successful sequences follow a path qualitatively very similar to the specific example described in detail here. Adopting a M33 distance of 840$\pm$20$\,$kpc, the sequences that match all observed properties within 1$\sigma$ errors, are constrained to host 96-99$\,$M$_\odot$ primaries, and 32$\,$M$_\odot$ (within 1$\,$M$_\odot$ uncertainty) secondaries in orbits with initial periods of 2.8-2.9$\,$days. If the full distance range of 750-1017$\,$kpc is adopted,  the progenitors are constrained to host primaries between 85-99$\,$M$_\odot$, secondaries between 28-32$\,$M$_\odot$, and initial orbital periods between 2.8-3.1$\,$days. Our model excludes secondaries at present more massive than $\simeq$65$\,$M$_\odot$ because they fail to explain the observed luminosity (see Figure \ref{luminosities}). Two factors contribute to the apparent underluminosity of the stellar component (see appendix B). On one hand, the orientation of the system with respect to our line of sight and associated projection effects reduce the star's measured luminosity ($\simeq$87\%); on the other hand the secondary  was not  born as a $\simeq$63-65$\,$M$_\odot$ star, but instead accreted much of its mass from the BH progenitor ($\simeq$13\%).  Looking at the system's properties at primary's collapse (Figure~\ref{SNparams}), our models show no ejection of baryonic mass. Furthermore, the allowed eccentricities post-BH formation are small (0.012-0.026). However, given the lack of kinematic information, we can not exclude kicks as high as $\sim$850$\,$km/s. This apparent discrepancy is explained by the change in orbital inclination at BH formation, which increases with the kick magnitude. Hence, our model constrains the kick to point mostly orthogonal to the orbital plane, without significantly affecting the eccentricity. 
\section{conclusions}
For the first time, we present a quantitative theoretical model that succeeds in explaining all observed properties of M33 X-7. The massive BH and its massive, but underluminous companion in a tight, slightly eccentric orbit are all physically explained, given our current understanding of massive binary evolution. Our results further shed light on the effect of mass transfer on massive stars and their internal structure, the relation between BH and progenitor masses, and asymmetries associated with BH formation through massive-star core collapse.
\begin{theacknowledgments}
  This work was partially supported by NSF grants AST--0908930 and CAREER AST--0449558 to VK.  TF is supported by a Northwestern University Presidential Fellowship. Simulations were performed on the computing cluster {\tt Fugu} available to the Theoretical Astrophysics group at Northwestern and partially funded by NSF grant PHY--0619274 to VK.
\end{theacknowledgments}

\bibliographystyle{aipproc}
\bibliography{myBiblio}
\section*{Appendix A}
Using the code STARS we create 3-25$\, $M$_\odot$ He star models with the same input physics used for the binary evolution calculations. We evolve each model after the end of the core He burning phase and calculate the duration ($t_{\rm{He}}$) of this phase and the amount of mass lost ($\Delta M$) in stellar wind as a function of the initial He star mass ($M_{\rm{He, i}}$). From these models we derive $t\rm_{He} =0.323221 + 6.24256\cdot M\rm_{He, i} ^{-1.45762}$ where $t\rm_{He}$ is in Myr, 
$\Delta M = 0.282 - 0.28\cdot M\rm_{He, i} + 0.052 \cdot M\rm_{He, i}^2 - 0.00102\cdot M\rm_{He, i}^3$ if $M\rm_{He, i} \leq 17\,M_\odot$, and $\Delta M =- 23.5355 + 10.262 \cdot \ln(M\rm_{He, i})$ if $M\rm_{He, i} > 17\,M_\odot$, where $M\rm_{He, i}$ and $\Delta M$ are in M$_\odot$.

\section*{Appendix B}
\noindent\textbf{Correction to the luminosity due to the inclination of the system:}
The shape of the star in M33 X-7
is distorted by rotation and tides, which cause the equatorial
regions to be colder than the poles (Von
Zeipel theorem), and dimmer. 
Given that our stellar models do not incorporate these effects on the star's surface temperature ($T_{\rm eff}$) and luminosity ($L$), 
we use the ELC code to determine an appropriate correction to apply to our models, and we find that $T_{\rm eff} \rightarrow{T_{\rm eff} / 0.954}$, and $\log(L) \rightarrow{ \log (L) - 0.14}$.
The observed luminosity is lower than the true luminosity because we are looking  at the colder equatorial regions of the star.

\smallskip
\noindent\textbf{Correction to the luminosity due to the partial-rejuvenation of the secondary:} 
Previous studies on the effect of accretion onto massive MS stars have shown that accreting stars not always rejuvenate, and that the result of mass accretion might be a star with a chemical structure unlike that of a single star of the same mass \cite{BraunLanger1995}. To a large extent, this effect is controlled by the assumed semiconvective mixing efficiency which, in turn, depends on the criterion for convection used in the stellar model.
\citet{BraunLanger1995} used the Ledoux criterion and showed that the non-rejuvenated models appear to be underluminous for their new mass during the remaining Main-Sequence evolution.
Following this idea, we use the Ledoux criterion in the single and binary evolution calculations, after having calibrated the semiconvective efficiency parameter against some of the results reported by \citet{BraunLanger1995}. 
Our results confirm that the rejuvenation of the secondary star after mass transfer is at most partial.

\end{document}